# SELF-SUPERVISED SPEECH REPRESENTATION AND CONTEXTUAL TEXT EMBEDDING FOR MATCH-MISMATCH CLASSIFICATION WITH EEG RECORDING

*Bo Wang[1,3], Xiran Xu[1,3], Zechen Zhang[1,3], Haolin Zhu[1,3], YuJie Yan[2,3], Xihong Wu[1,3], Jing Chen[1,2,3]*

[1]Speech and Hearing Research Center, School of Intelligence Science and Technology, Peking University
[2]National Biomedical Imaging Center, College of Future Technology, Peking University
[3]National Key Laboratory of General Artificial Intelligence, China
janechenjing@pku.edu.cn

## ABSTRACT

Relating speech to EEG holds considerable importance but is challenging. In this study, a deep convolutional network was employed to extract spatiotemporal features from EEG data. Self-supervised speech representation and contextual text embedding were used as speech features. Contrastive learning was used to relate EEG features to speech features. The experimental results demonstrate the benefits of using self-supervised speech representation and contextual text embedding. Through feature fusion and model ensemble, an accuracy of 60.29% was achieved, and the performance was ranked as No.2 in Task 1 of the Auditory EEG Challenge (ICASSP 2024). The code to implement our work is available on Github: https://github.com/bobwangPKU/EEG-Stimulus-Match-Mismatch.

*Index Terms*— Auditory EEG decoding, self-supervised speech representation, contextual text embedding

## 1. INTRODUCTION

Match-mismatch task in ICASSP 2024 Auditory EEG Challenge is a classification problem. Given an EEG segment and five stimulus segments, the task was to classify which one of them corresponds to the EEG. In the previous work, low-level features (i.e., envelope, mel-spectrograms) were extracted from the speech, and then these extracted features, along with the EEG segment, were transformed into a latent space using neural network, aiming to maximize the similarity between EEG signals and the features of the target stimuli [1][2].

Two main factors for those work may potentially limit model performance. Firstly, only low-level features of speech were used. Recent studies have revealed that neural responses also encode high-level features (i.e., phonetic, lexical, syntax, and semantic) [3]. Specifically, speech representations extracted from self-supervised speech models (e.g., wav2vec) or text embeddings derived from pre-trained language models (e.g., GPT) were both helpful for aligning brain signals and stimuli [4][5]. Second, the neural networks used were relatively shallow, imposing constraints on the effectively EEG feature extracting.

To address these limitations, a deep convolutional network with residual connections was used to extract spatiotemporal features from EEG data. Besides, self-supervised speech representation and contextual text embedding were utilized as speech features. Furthermore, a contrastive learning framework was implemented to enhance the discrimination between matched and mismatched stimuli.

This work was supported by the National Key Research and Development Program of China (No.2021ZD0201503), a National Natural Science Foundation of China (No.12074012), and the High-performance Computing Platform of Peking University.

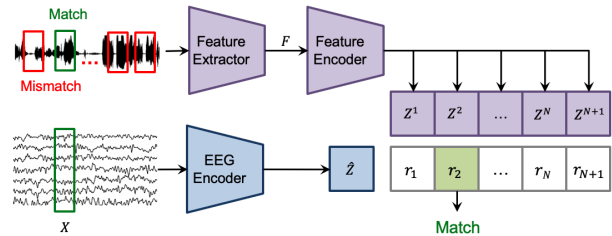

**Fig. 1** Contrastive learning training framework.

## 2. METHODS

### 2.1 Model

The framework of the proposed model is shown in Fig. 1. For a segment of EEG data $X$, the matched and $N$ mismatched speech segments are fed into a feature extractor to get the speech features $F$. These mismatched speech segments are randomly sampled from other segments of the same stimulus. An EEG encoder and a feature encoder are used to encode EEG data and speech features into a latent space, resulting $\hat{Z}$ and $Z^i$ ($i = 1, 2, \ldots N + 1$) respectively. Both $\hat{Z}$ and $Z^i$ comprise $D$ time series, i.e., $Z^i = (z_1^i, z_2^i, \ldots, z_D^i)^T$. Here, $D$ represents dimensionality of the latent space, and $T$ donates transpose. The InfoNCE loss is used in this work:

$$L = -\sum_{d=1}^{D} \log \frac{\exp\left(\mathrm{sim}(\widehat{z_d}, z_d^p)\right)}{\sum_{i=1}^{N+1} \exp\left(\mathrm{sim}(\widehat{z_d}, z_d^i)\right)} \quad (1)$$

where $\mathrm{sim}(\cdot,\cdot)$ represents Pearson correlation between two time series, and p is the index of the matched segment in $Z^i$.

The architecture of the EEG encoder is demonstrated in Fig. 2. EEG $X$ is first fed in a linear layer and followed by a $1 \times 1$ convolution layer with channel $D_h$. Then, a stack of five blocks of three convolutional layers is applied to extract EEG spatial-temporal features. The convolutional layer is the same as that in previous work [4]. Finally, the convolutional blocks output is sequentially fed into a $1 \times 1$ convolution layer with $2D_h$ channels, followed by a GELU activation, and then another $1 \times 1$ convolution layer with $D$ channels to get the latent representation $\hat{Z}$. For the speech feature, $F$ was fed into a $1 \times 1$ convolution layer with $2D$ channels, followed by a GELU activation, and then another $1 \times 1$ convolution layer with $D$ channels to get the latent representation $Z$.

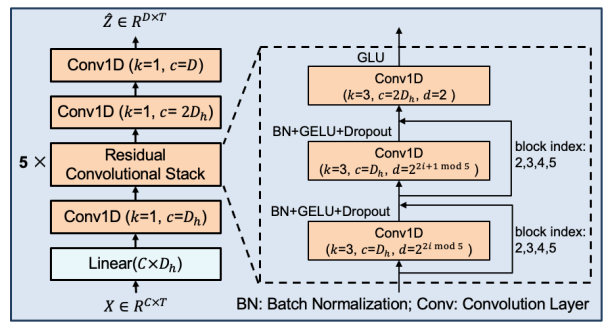

**Fig. 2** The architecture of the EEG encoder.

## 2.2 Dataset

We utilized the preprocessed EEG data provided by the challenge organizers, which we referred to as broadband EEG [6]. Additionally, a filter bank (0-4 Hz, 4-8 Hz, 8-12 Hz, and 12-30 Hz) was employed to perform bandpass filtering on the broadband EEG data, resulting in multi-band EEG data. Five-fold cross-validation with an "unseen subjects & unseen stimuli" data partitioning strategy was used. Specifically, the data from a portion of subjects were assigned as validation set (subjects in the validation set for the five folds: 1-26, 18-34, 35-51, 52-68, 69-85), while the data from the remaining subjects were allocated to the training set. For the validation set, we excluded the EEG recordings that corresponded to the stimuli encountered in the training set.

## 2.3 Speech feature extraction

Self-supervised speech model wav2vec 2.0 was used to extract speech representation. In practice, we used the wav2vec2-large-xlsr-53-dutch[1]. During the process, speech segments of 5 seconds were provided as the input to the model. The output of the model's 14th layers was utilized as speech representation. The representation initially had a dimensionality of 1024, which was then reduced to 64 dimensions using PCA (Principal Component Analysis).

Pretrained language model GPT-2 was used to extract contextual text embedding. In practice, we used the gpt2-small-dutch[2]. Stabilizing timestamps for whisper was used to get text and word temporal information (i.e., onset and offset time) of the speech[3]. During the process, each word along with its preceding 4 words was fed into GPT-2. The 9th layer's output was used as word contextual embedding. The embedding initially had a dimensionality of 768, which was then reduced to 4 dimensions using PCA. To create continuous text embedding, a multivariate time series was constructed with each word embedding filling in its corresponding time slot. More details can be found in [5].

To match the sampling frequency of the EEG data, the wav2vc speech representations were resampled to 64 Hz and the continuous word embedding was set to 64 Hz. The continuous word embedding was also 4 Hz low-pass filtered according to our pilot study.

## 2.4 Experiment setup

The EEG data and the corresponding speech features were split into segments with a duration of 5 seconds without overlap, and then were standardized across time. During training, an Adam optimizer with a learning rate of $2e^{-5}$ was used. The batch size was set to 32, and the unmatched speech segment number $N$ was set to 32. The hidden size $D_h$ was set to 256. The dropout rate was set to 0.5. For validation, 4 unmatched segments were sampled. The predicted matched segment was determined by selecting the speech segment that had the highest correlation with the EEG segment in the latent space among all the speech segments. The classification accuracy was evaluated on the validation set for every 1k steps. The training was stopped when the classification accuracy was not increased for 20 consecutive evaluations.

## 4. RESULT

**Tab. 1** Classification accuracy on the validation set with different speech feature (env: envelope, mel: mel-spectrogram, wav2vec: wav2vec 2.0 representation, gpt: GPT text embedding) and different filtering EEG. The accuracy was averaged over 5 folds. Standard deviation was calculated across 5 folds.

| Speech Features | Classification Accuracy (%) | |
|---|---|---|
| | broadband EEG | multi-band EEG |
| env | 43.34±1.44 | 44.10±1.81 |
| mel | 50.51±2.51 | 50.45±2.20 |
| env+mel | 54.67±2.02 | 54.92±2.22 |
| wav2vec | 69.95±2.60 | 70.21±2.22 |
| gpt | 34.84±2.37 | 36.72±1.43 |
| env+mel+ wav2vec | 71.04±2.51 | 71.63±2.33 |
| env+mel+ wav2vec +gpt | 71.54±2.79 | 71.62±1.98 |

The classification accuracy on the validation set is shown in Table 1. The results show that both self-supervised speech representation and contextual text embedding are effective in aligning EEG and speech. Among these features, the wav2vec representation stands out as the most effective. Previous studies have shown that the wav2vec representation encodes phoneme- and word-level information besides spectrotemporal features[7]. Therefore, these results suggest the importance of high-level features of the speech in improving the model's performance. Furthermore, integrating multiple features and replacing the broadband EEG with multi-band EEG further enhances the performance.

Similar to [2], we also utilize ensemble learning to enhance model performance. A total of 136 models were trained by combining different filtering EEG and various speech feature combinations. The classification output was determined through voting. An accuracy of 60.29% was achieved on the test set of the challenge. As this performance was significantly worse than the result reported on the local dataset, we suspect the presence of overfitting. Further experiments should be done when the online test set was available.

## 5. CONCLUSION

In this work, we introduce self-supervised speech representation and contextual text embedding into match-mismatch classification of speech and EEG recording. Experimental results show that both types of features are effective in relating EEG and speech. When combined with low-level features, the model's performance is further enhanced.

## 6. REFERENCE

[1] M. Borsdorf, S. Pahuja, G. Ivucic, S. Cai, H. Li, and T. Schultz, "Multi-head attention and GRU for improved match-mismatch classification of speech stimulus and EEG response," in *Proc. ICASSP*, 2023.
[2] M. Thornton, D. Mandic, and T. Reichenbach, "Relating EEG recordings to speech using envelope tracking and the speech-FFR," in *Proc. ICASSP*, 2023.
[3] C. Puffay, J. Vanthornhout, B. Accou, H. Van Hamme, and T. Francart, "Robust neural tracking of linguistic speech representations using a convolutional neural network," *J Neural Eng*, vol. 20, no. 4, 2023.
[4] A. Défossez, C. Caucheteux, J. Rapin, O. Kabeli, and J.-R. King, "Decoding speech perception from non-invasive brain recordings," *Nat Mach Intell*, vol. 5, pp. 1097-1103, 2023.
[5] B. Wang, X. Xu, L. Zhang, B. Xiao, X. Wu, and J. Chen, "Semantic reconstruction of continuous language from MEG signals." *arXiv preprint*, arXiv:2309.07701, 2023.
[6] B. Accou, L. Bollens, M. Gillis, W. Verheijen, H. V. Hamme, and T. Francart, "SparrKULee: A speech-evoked auditory response repository of the KU Leuven, containing EEG of 85 participants." bioRxiv, p. 2023.07.24.550310, 2023
[7] A. R. Vaidya, S. Jain, and A. G. Huth, "Self-supervised models of audio effectively explain human cortical responses to speech." in *Proc. ICML*, 2022.

---

[1] https://huggingface.co/jonatasgrosman/wav2vec2-large-xlsr-53-dutch
[2] https://huggingface.co/GroNLP/gpt2-small-dutch
[3] https://github.com/jianfch/stable-ts